\documentclass[aps,prl,twocolumn]{revtex4-1}

\usepackage{graphics}      
\usepackage{graphicx}      
\usepackage{longtable}     
\usepackage{url}           
\usepackage{bm,color}            
\usepackage{amssymb, amsmath, amsthm, enumerate, epsfig,subfigure}
\usepackage{multirow}

\theoremstyle{plain}
\newtheorem*{prop*}{Proposition}

\definecolor{red}{rgb}{0.8, 0.25, 0.33}
\definecolor{green}{rgb}{0.0, 0.5, 0.0}

\begin{document}

\title{An Ising machine based on networks of subharmonic electrical resonators}

\author{L. Q. English$^1$, A. V. Zampetaki$^2$, K. P. Kalinin$^3$, N. G. Berloff$^3$, P. G. Kevrekidis$^4$}
\affiliation{$^1$Department of Physics and Astronomy, Dickinson College, Carlisle, Pennsylvania 17013, USA \\
$^2$Institut f\"{u}r Theoretische Physik II, Weiche Materie, Heinrich-Heine-Universit\"{a}t, 40225 D\"{u}sseldorf, Germany \\
$^3$Department of Applied Mathematics and Theoretical Physics, University of Cambridge, CB3 0WA, UK \\
$^4$Department of Mathematics and Statistics, University of Massachusetts, Amherst, Massachusetts 01003-4515, USA}

\date{\today}

\begin{abstract}
  We explore a case example of networks of classical electronic oscillators evolving towards the solution of complex optimization problems. We show that when driven into subharmonic response, a network of such nonlinear electrical resonators can minimize the Ising Hamiltonian on non-trivial graphs such as antiferromagnetically coupled rewired-M{\"o}bius ladders. In this context, the spin-up and spin-down states of the Ising
  machine are represented by the oscillators' response at the even or odd
  driving cycles. Our experimental setting of driven nonlinear oscillators coupled via a programmable switch matrix leads to a unique energy minimizer when one
  such exists, and probes frustration where appropriate.
  Theoretical modeling of the electronic oscillators and their couplings allows
  us to accurately reproduce the qualitative features of the experimental results.
 This suggests the promise of this setup as a  prototypical one for exploring the
  capabilities and limitations of such an unconventional computing platform.
\end{abstract}

\maketitle

\section*{Introduction}

The desire to solve complex combinatorial problems in energy and time-efficient manner ignites the race to implement classical state-of-the-art optimisation techniques on traditional hardware. The implementation of the simulated annealing on CPU leads to a traditional classical solver, on complementary metaloxide-semiconductors (CMOSs) hardware results in the CMOS annealer \cite{yamaoka201520k,takemoto20192}, and with field programmable arrays (FPGAs) it
is known as the digital annealer machine  \cite{tsukamoto2017accelerator,matsubara2020digital}. The realisation of another physics-inspired method on GPUs underlies the simulated bifurcation machine \cite{goto2019combinatorial,goto2021high}. With such mature dedicated hardware, the computational performance of classical optimisation methods can be studied on a large scale of hundreds of thousands of elements.

Novel computing paradigms can be based on novel physical platforms augmented by traditional hardware. In such a hybrid approach, the optimisation efficiency depends not only on classical algorithms, and the better quality of solutions is expected from natural internal processes in physical systems, while the classical hardware provides interactions between physical elements. For example, the FPGA operates in concert with the optical parametric oscillators \cite{McMahon2016} and the spatial light modulator can create couplings between polariton condensates \cite{Berloff2017,Kalinin2020} for solving hard optimisation problems.

To overcome the time limitations of traditional hardware, the pure passive unconventional computing architectures can be considered. In these architectures, the solution to the optimisation problem is found solely through an analogue system without exchanging information with the classical counterparts.
The memristors (short for memory resistors) can perform matrix-vector multiplications according to Ohm's and Kirchhoff's laws in a completely analogue way \cite{strukov2008missing}. Circuits of memristors (memristor crossbars) are used for simulating neural networks \cite{hu2018memristor,yao2020fully,lin2020three} including Hopfield networks for solving hard optimisation problems \cite{cai2020power}. A further improvement in power consumption over memristor-based Hopfield networks is expected for networks of phase-transition nano-oscillators with capacitive couplings \cite{dutta2020ising}. These beyond-traditional hardware approaches \cite{yang2013memristive,li2018review,kalinin2020nonlinear}, as well as all-optical passive computing architectures with a similar principle of in-memory computing \cite{RoquesCarmes2020,shen2017deep,prabhu2020accelerating,bernstein2020freely}, are naturally suitable for highly parallel calculations and offer orders-of-magnitude higher energy efficiency than classical devices.
Many more physical systems are under intense investigation as quadratic unconstrained binary optimisation (QUBO) solvers in the post-CMOS era including lasers \cite{Babaeian2019,Pal2019,Parto2020,gershenzon2020exact}, photonic simulators \cite{Pierangeli2019}, trapped ions \cite{kim2010quantum}, photon and polariton condensates \cite{klaers2010bose,Kassenberg2020}, QED \cite{guo2019sign,marsh2020enhancing}, and others \cite{Boehm2019,okawachi2020demonstration,cen2020microwave}.

The electronic and optical oscillator-based unconventional computing machines are generally applied to the minimization of spin Hamiltonians, to which many of the real-life optimisation problems can be mapped with a polynomial overhead \cite{lucas2014ising}. One of the challenges in assessing the potential optimisation performance of such platforms is caused by the choice of instances of NP-hard problems. For example, minimising the Ising spin Hamiltonian on unweighted 3-regular graphs is proven to be NP-hard \cite{Garey1974}, while for a subclass of M{\"o}bius ladder graphs, which are often chosen for testing non-traditional computing platforms \cite{McMahon2016,Takata2016,Yamamoto2017,Boehm2019,Chou2019,cen2020microwave,okawachi2020demonstration,dutta2020ising}, the Ising model can be minimized in polynomial time \cite{kalinin2020complexity}.

To develop new physics-inspired algorithms and explore non-trivial ways for escaping local minima of complex optimisation problems, the easy-to-assemble circuits of electronic oscillators could be considered. Although this is a well-studied classical system, there are only a mere handful of works with physical implementations of oscillator-based circuits, with most studies devoted to theoretical and numerical simplified models \cite{xiao2019optoelectronics}, which do not necessarily represent internal physical processes that can be critical to optimisation performance. There exist many types of electrical oscillators one may use for computing. The vertex colouring problem of unweighted graphs has been recently addressed with small networks of five coupled relaxation oscillators with capacitive connections \cite{parihar2017vertex}. An integrated circuit of 30 relaxation oscillators with programmable couplings was implemented for solving the maximum independent set problem \cite{mallick2020using}. The all-electronic Ising Machine has been explored with weighted resistive couplings for four CMOS LC oscillators \cite{Chou2019} with larger network of 240 oscillators implemented on a chimera-graph architecture \cite{wang2019new}.

In this work, we explore possible global optimisation mechanisms that could help to evaluate the new small-size physical solvers by minimising the Ising Hamiltonian with fundamental passive electrical circuit elements: the resistor, the capacitor and the inductor, in the presence of nonlinearity. The electrical network of such RLC oscillators is an example of a purely classical computing system implemented on CMOS. For such electronic oscillator networks, we show the difference between the Ising minimization of the trivial problems, such as Max 3-regular cut on the M{\"o}bius ladder graphs, and the non-trivial, such as on the rewired M{\"o}bius ladder graphs and on random 3-regular graphs. The ground state success probability for non-trivial problems can be dramatically increased using the dynamic control of the inductance, the optimal value of which helps to efficiently escape the local minima. We discuss possible ways for creating easy-reconfigurable couplings between oscillators and possibilities for the large-scale on-chip integration of electronic circuits. Better energy-efficiency could be further achieved with networks of energy-recycling electronic oscillators, in which the energy is converted between two forms, electrostatic and magnetic energies, during each oscillation cycle. Such conventional integrated electronic circuits could not offer better power consumption than passive optical computing architectures but rather can open opportunities to study non-trivial ways for escaping local minima and facilitate the development of physics-inspired algorithms. The behaviour of electronic oscillators for solving hard problems may be further generalised to the synchronisation dynamics of coupled nonlinear oscillators of different nature.

\section{Experimental Setup}
The basic idea is to drive a collection of nonlinear oscillators at a frequency that is roughly twice their natural frequency, $\omega_d\approx 2\omega_0$, such that subharmonic resonance is induced in them (see also \cite{english2012}). Subharmonic resonance is a nonlinear phenomenon and (in the case of an isolated oscillator) its onset occurs above a  threshold amplitude in the driving signal~\cite{nayfeh}. It is characterized by an oscillator response that repeats every other driver period. Therefore, two response states are conceivable \cite{cai2020power}, namely an oscillator response corresponding to either {\it even} or {\it odd} driving cycles. These two oscillator states will represent the basic ``spin-up'' and ``spin-down'' states of the Ising machine.

\begin{figure}
\includegraphics[width=0.45\textwidth]{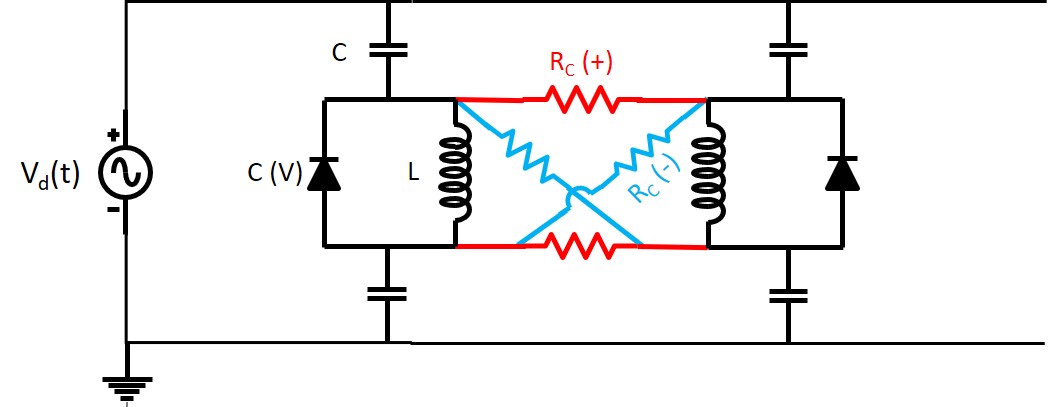}
\caption{The main idea of coupling between nodes is illustrated here using a pair of oscillators, each consisting of a varactor diode, with capacitance C(V), and an inductor, L, in parallel. These are driven via capacitors, C, to induce subharmonic oscillations, for which $\omega = \omega_d /2$. The two oscillators can be either positively coupled using the resistor pair labeled $R_c(+)$ (red), or negatively using the resistor pair labeled $R_c(-)$ (blue). The oscillations across each oscillator's diode/inductor are measured as a floating voltage.}
\label{osc}
\end{figure}

While the earlier work of~\cite{cai2020power} proposed generic nonlinear oscillators driven by dedicated noise generators to induce parametric resonance, this is not feasible with the nonlinear RLC oscillators used here. Instead, we employ a single sinusoidal voltage signal (from a function generator) to drive all oscillators via capacitors into subharmonic resonance, as shown in Fig.~\ref{osc}. The oscillator consists of a varactor diode (NTE 618), featuring a nonlinear dependence
of the capacitance on the voltage C(V), and an inductor, L. The coupling between a given pair of oscillators is achieved via resistors. Resistors connected straight across (red, labeled $R_c(+)$) favor in-phase oscillation between the two oscillators, whereas crossed resistors (blue, labeled $R_c(-)$) favor out-of-phase oscillation. The measured resistances of $R_c$-resistors were the same to within 1\%, the inductor values to within 0.25\%, and the capacitors to within 1\%.

\begin{figure}
\includegraphics[width=0.4\textwidth]{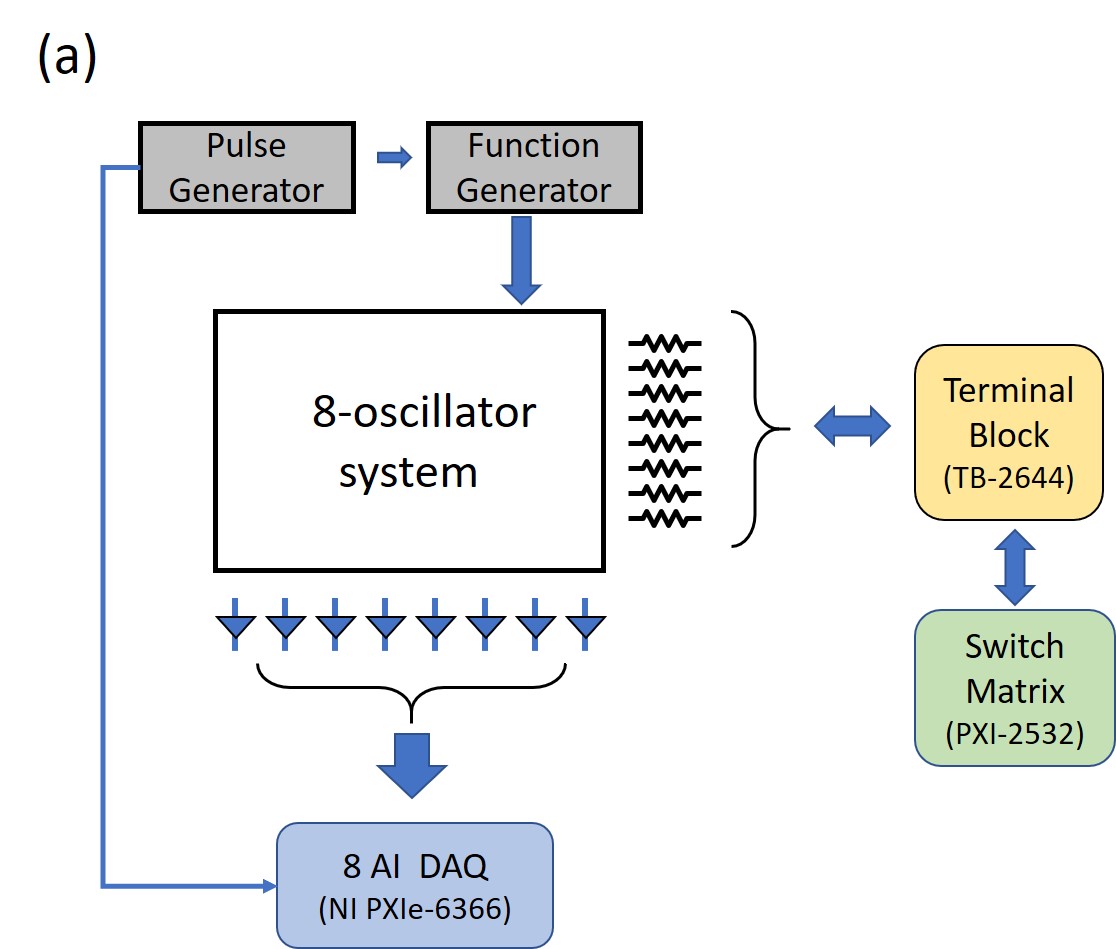} \\
\includegraphics[width=0.45\textwidth]{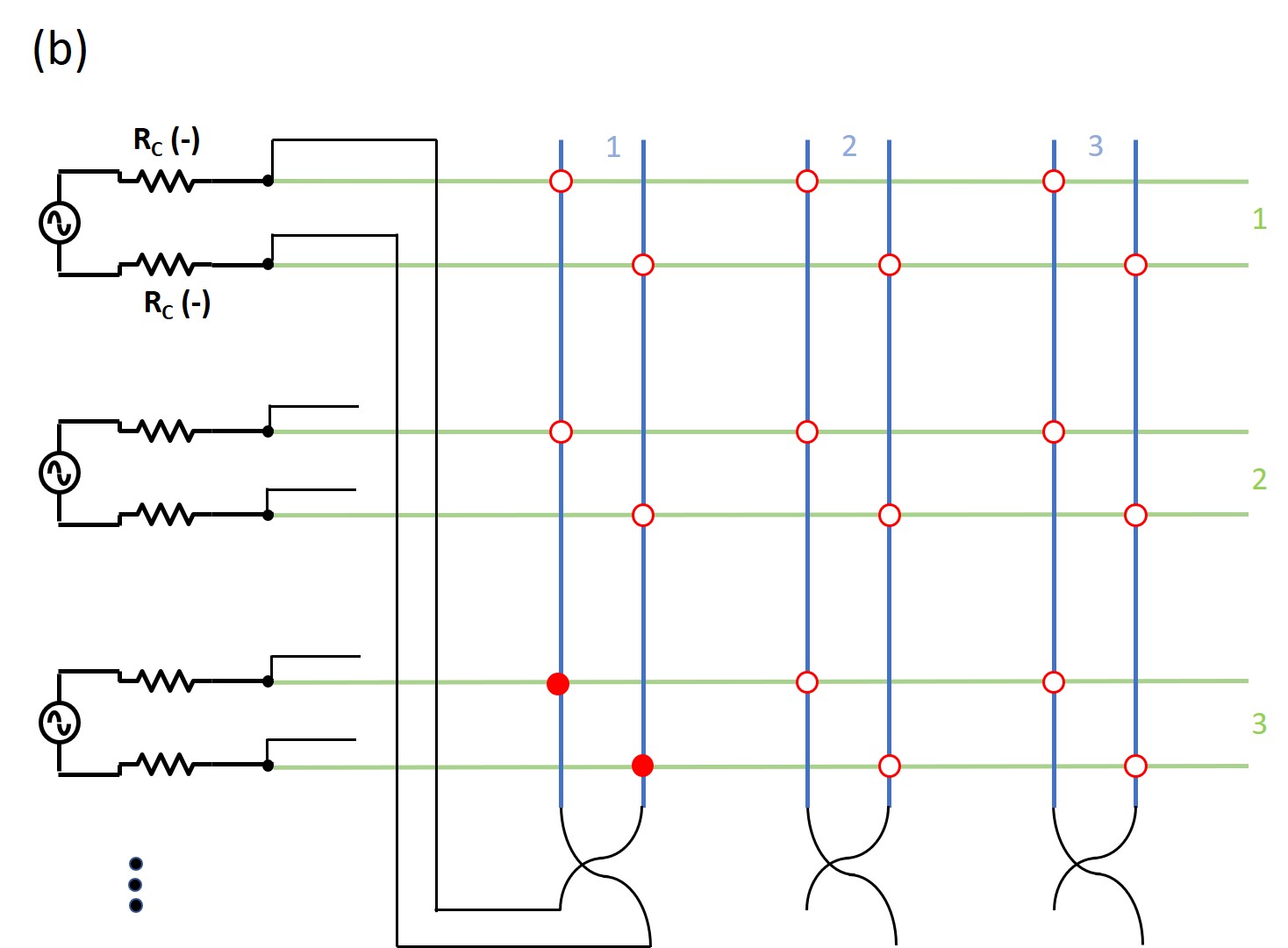}
\caption{(a) Schematic of the basic experimental setup: the oscillator-system is driven sinusoidally, starting at the trigger of the pulse generator, which also initiates the data collection at the DAQ board. The oscillators are impedance-isolated from the AI channels of the DAQ via buffers. The coupling network connecting the oscillators is established via a terminal block (TB-2644) set to the 2-wire 8-32 configuration, and the switch-matrix unit (PXI-2532).
(b) A more detailed view of the coupling network using the switch matrix: there are 8 inputs and 32 outputs in this configuration (16 of which are used). The inputs connect directly to the oscillators. The first 8 outputs connect back to the inputs in a one-to-one fashion, but the second 8 outputs shown in this figure cross the two wires before connecting back, as shown. The electrical switches at the cross-points of this switch-matrix module, represented here by open and closed red circles, can be programmed to be open or closed. As shown here, oscillator 1 and 3 would be negatively coupled.}
\label{exp_setup}
\end{figure}

Figure \ref{exp_setup}(a) schematically depicts the experimental system for a network of 8 subharmonic resonators. The main experimental challenge is to connect these 8 oscillators via a programmable and reconfigurable coupling network. Our solution was to use a switch-matrix module that can be configured (via the terminal block) into a two-wire 8x32 cross-point matrix. The 8 analog-in channels of a data acquisition card (NI PXIe-6366) are synchronized to the start of the driving signal via a pulse generator and digitize the voltage profiles at all 8 oscillators.
The coupling scheme is illustrated in greater detail in Fig.\ref{exp_setup}(b), which shows the example of a negative coupling between oscillator 1 and 3. There are eight inputs to the module arranged vertically on the left (three of which are depicted), and 16 used outputs arranged horizontally at the bottom (again three are shown). The first 8 outputs (not shown) are responsible for positive coupling between oscillator pairs, and the next 8 outputs (three shown) are responsible for the negative coupling. The latter is accomplished by crossing the wire at the bottom before feeding it back in to the respective oscillator. By closing that particular switch-pair (see solid circles), oscillators 1 and 3 are connected in the same manner as represented by the blue resistors in Fig.\ref{osc}.

\section{The Model}
As was shown in Ref. \cite{faustino2011}, we can model the varactor diode, the nonlinear circuit element, as a parallel combination of three idealized components: a nonlinear capacitor of variable capacitance, C(V), a nonlinear resistor whose current-voltage relationship is given by $I_D(V)$, and a nonlinear dissipation resistance, $R_l$. We then apply the Kirchhoff loop rule using two loops around the circuit shown in Fig.~\ref{osc}, while also keeping mathematical track of the currents entering the $n$th node through the top capacitor and exiting through the lower capacitor. 
The detailed steps in the analysis are relegated to Appendix A; here we show only the final set of non-dimensionalized equations of motion governing this electrical network
that will be used for the simulation results presented below. More specifically,
the voltage dynamics for each oscillator (indexed by $n$) reads:

\begin{align}
\label{govern}
&\left[1 + 2 c(v_n) \right] \frac{dv_n}{d\tau} =\Omega \cos(\Omega \tau) - \frac{2}{\tau_c} \left(\frac{R_c}{R_l}\right) v_n + \\ \nonumber
& 2\left[i_D(v_n)-y_n\right]-\frac{1}{\tau_c}\sum_m B_{nm} (v_n+v_m); \\ \nonumber
&\frac{dy_n}{d\tau}=v_n ,
\end{align}
where the symbols are defined as follows in terms of the measurable circuit quantities:
$v_n=V_n/A$, with $A$ being the amplitude of the driving signal and $V_n$ the voltage across the diode; $y_n=Y_n/(A C_d \omega_0)$, with $Y$ representing the current through the inductor. Similarly, $i_D=I_D/(A C_d \omega_0)$, where $I_D(V)$ is the voltage-dependent current through the varactor diode. $C(V)$ is the voltage-dependent capacitance of the diode, and $c=C(V)/C_d$. (Both functions, $I_D$ and $C$, are given in the appendix.) Furthermore, $\omega_0 = 1/ \sqrt{L C_d}$ and
$\tau=\omega_0 t$, and $\Omega=\omega/\omega_0$ represent the
adimensionalized time and driving frequency. Finally, $\tau_c = R_c C_d \omega_0$, and $B_{nm}$ is either zero (no connection between that node pair) or 1  when the pair is negatively coupled.

\section{Experimental Results}
\begin{figure}
\includegraphics[width=0.45\textwidth]{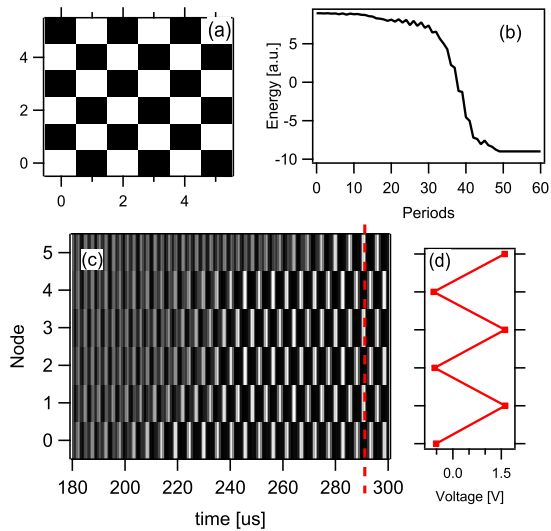}
\caption{(a) The coupling matrix for N=6 M{\"o}bius-ladder graph. (b) the time evolution of the system's energy, settling at the ground-state energy of $E=-8$ after around 50 subharmonic periods. (c) the N=6 circuit response - time is plotted on the horizontal axis, node number vertically, and the voltage response is depicted  in gray-scale. (d) the voltage profile encoding the ground state at particular instant of time, depicted as the red, dashed line in (c).}
\label{data0}
\end{figure}

Let us begin by examining an antiferromagnetically coupled M{\"o}bius ladder graph for N=6. The idea is to minimize the Ising Hamiltonian, which means finding the spin configuration $\{s_i=\pm1\}$ that yields the minimum energy for $E_{Ising}=-\frac{1}{2}\Sigma_{i,j}J_{ij} s_i s_j$. Solving the Max 3-regular cut problem on an unweighted graph is trivially formulated as minimizing the Ising Hamiltonian by assigning $J_{ij}=-1$ to connecting edges.  This coupling network is shown schematically in Fig.~\ref{data0}(a), where black (white) squares represent negative (zero) coupling between that pair of nodes. It is straightforward to see that this network that has a unique lowest-energy solution (up to a minus sign) of $[1, -1, 1, -1, 1, -1]$. When we drive the lattice with this coupling network at $f=380$ kHz and $V_d=4$ V, the system is driven to that lowest energy ($E=-9$), as seen in Fig.~\ref{data0}(b), and we get the voltage response depicted in gray-scale in Fig.~\ref{data0}(c). We see that after about $280 \mu$s, or 50 subharmonic periods, the final alternating pattern firmly establishes itself. A time snapshot of the voltage profile across all six nodes - at a time indicated by the red dotted line in panel (b) - is shown in Fig.~\ref{data0}(d). It is evident that the correct solution is encoded in that voltage profile. It should be mentioned that we computed the configurational energy from experimental data as outlined in Appendix B.

Note that for this network, there is no frustration, the optimization state is unique, and the electrical circuitry ``finds'' this state quickly and with complete reliability.
This is true for any network that admits a single optimal solution without frustration. In such cases (i.e., for M{\"o}bius graphs when N/2 is odd), the circuit was found to perform with perfect accuracy. To demonstrate practical use for computing, however, the system also has to find solutions for the larger class of networks with frustration. In the frustrated ground state, some spins would have to be aligned in spite of being coupled antiferromagnetically. This would happen if N/2 is even in M{\"o}bius ladder graphs. 

\begin{figure}
\includegraphics[width=0.5\textwidth]{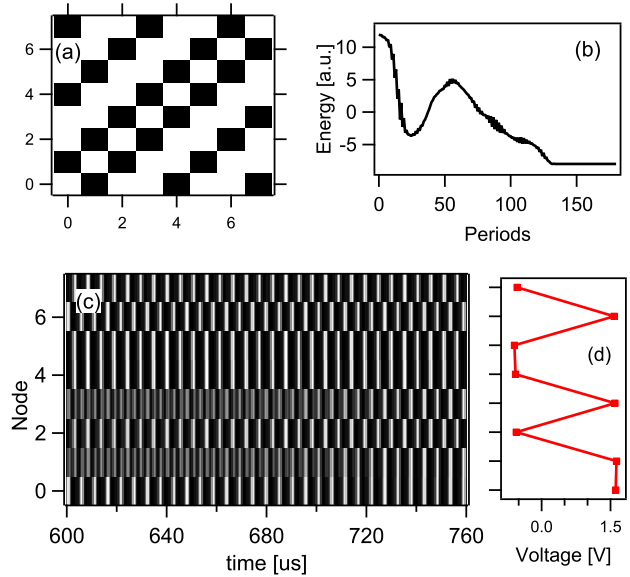}
\caption{(a) The coupling matrix for N=8 M{\"o}bius-ladder graph. (b) the time evolution of the system's energy, settling at the ground-state energy of $E=-8$ after around 130 subharmonic periods; interestingly the system arrives there non-monotonically. (c) the circuit response encoding the solution of lowest energy, depicted as the temporal voltage snapshot in (d).}
\label{data0b}
\end{figure}

Let us now examine the N=8 M{\"o}bius ladder, depicted in Fig.~\ref{data0b}(a). The experimental results for this network are displayed in Fig.~\ref{data0b}(b)-(d). Panel (b) shows the energy evolution of the state, as computed from Eq.~(\ref{ising}). The energy does eventually reach the lowest possible value for this network (after about 130 subharmonic periods), but it does not reach it monotonically. Panel (c) and (d) reveal that the oscillator final response pattern encodes the state $[1,1,-1,1,-1,-1,1,-1]$, which is one of the degenerate ground states with an energy of $E=-8$. Note that this network does exhibit frustration - for instance, nodes 0 and 1 are negatively coupled, but this optimal state has those same two nodes oscillate in synchrony.

\begin{figure}
\includegraphics[width=0.48\textwidth]{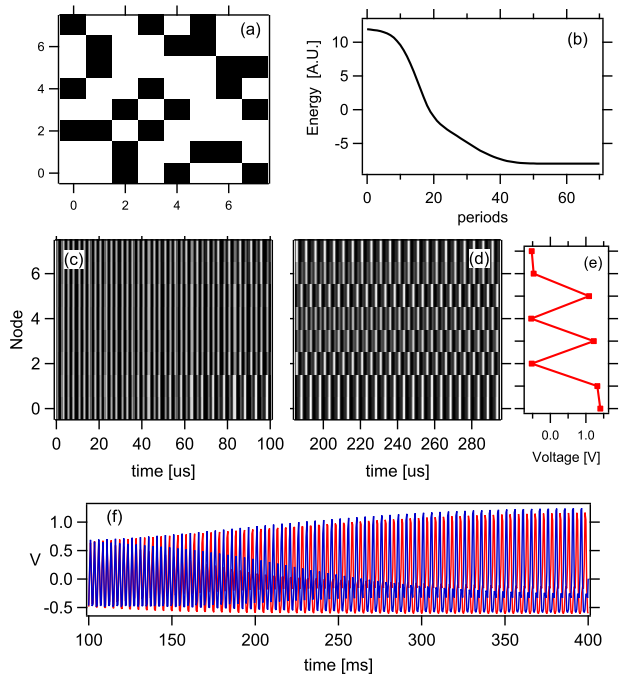}
\caption{(a) Another 3-regular graph of N=8. (b) energy evolution of the network; we reach a ground state of energy $E=-8$ after around 45 subharmonic periods. (c), (d) early and late circuit response, respectively. (e) the final state, as encoded in the voltage profile. (f) time-evolution of the voltage at node 3 (red) and 4 (blue).}
\label{data1}
\end{figure}

Figure \ref{data1} relates to a different 3-regular graph - comparing Figs.~\ref{data0b}(a) and \ref{data1}(a) reveals the coupling modifications. The raw data is shown (in the manner of previous figures) in panels (c) and (d), which depict the initial and final time-interval responses. Figure \ref{data1}(b) computes the configurational energy, according to Eq.~(\ref{ising}) (in the Appendix) as before, at each period of oscillation. It is evident that after around 50 subharmonic periods (or about 250 $\mu$s), the electronic system has settled into the final state of the minimum energy, $E=-8$. The panels (b)-(d) also illustrate that in the evolution towards the final state,
certain parts of the eventual state emerge much earlier than others. In this example, nodes 0 and 1 come into synchrony early, at around 70 $\mu$s, whereas nodes 3 and 4 do not snap into an anti-synchronous response until late, between 200 and 300 $\mu$s. This is illustrated in panel (f), which plots the voltage profiles of nodes 3 and 4 (red and blue trace, respectively). 

\begin{figure}
\includegraphics[width=0.45\textwidth]{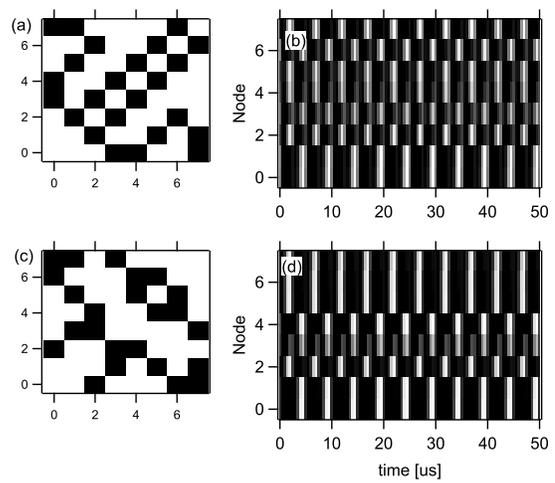}
\caption{Two additional 3-regular graphs, depicted in (a) and (c). Panels (b) and (d) show the steady-state circuit response yielding the respective ground states of energy $E=-8$.}
\label{data2}
\end{figure}

As two final examples of 3-regular graphs, consider Fig.~\ref{data2}(a) and (c). The driver frequency is again 400 kHz, the driver amplitude is gradually raised until a subharmonic pattern first emerges, and the steady-state circuit responses are shown in Fig~\ref{data2}(b) and (d), respectively. Both states encoded here in the voltage pattern match one of the optimized solutions for these graphs. For the two graphs they are, respectively, $[1,1,-1,1,-1,-1,1,-1]$ and $[1,1,-1,1,1,-1,-1,-1]$, both of which yield an energy of $E=-8$ for their respective networks.

It should be emphasized that these ground-state solutions in these 3-regular graphs compete with other patterns of fairly low energies, and such patterns can also emerge at or near the driver-amplitude threshold. In fact, when the driver is turned off and then on again, the same pattern does not always reappear even in the absence of changes to the driving conditions. A detailed statistical analysis has not been attempted yet but
would clearly be an interesting topic for further study.

Furthermore, in order to attain the ground states, in some cases it proved necessary to randomly permute the inductors for the eight oscillators. The measured inductance values for all inductors agreed to within 0.25\%, but even that low level of spatial ``noise'' in some instances proved sufficient to prevent the evolution to one of the correct ground states; here a mere rearranging of that noise would allow such states to manifest. In effect, our experimental results suggested the relevance of introducing some inductor noise to move the  system out of local minima and nudging it towards the global minimum.

\section{Numerical Simulations of Electrical Circuits}

We now turn to numerical simulations of this system described by Eq.~(\ref{govern}). Such simulations add three important facets to the picture: (i) they can, in
principle, be used to map out more systematically the role of noise, initial conditions, and driving parameters,  (ii) they allow us to more easily perform a statistical test, evaluating the efficiency of this computational scheme, and (iii) they allow an investigation of larger systems than can be currently implemented experimentally.

Our aim in this first proof-of-principle work is to reproduce in the simulations some of the experimental results shown previously. The numerical integration of Eq.~(\ref{govern}) leads quickly to the correct ground state  for networks without frustration. For instance, in the antiferromagnetically coupled ring with N=8, this happens within roughly 10 subharmonic periods, or around 50 $\mu$s. This time is shorter than what we see in Fig.~\ref{data0}, but with higher driving amplitudes the experimental time can be reduced to align more closely with the simulations. 

\begin{figure}
\includegraphics[width=0.98\columnwidth]{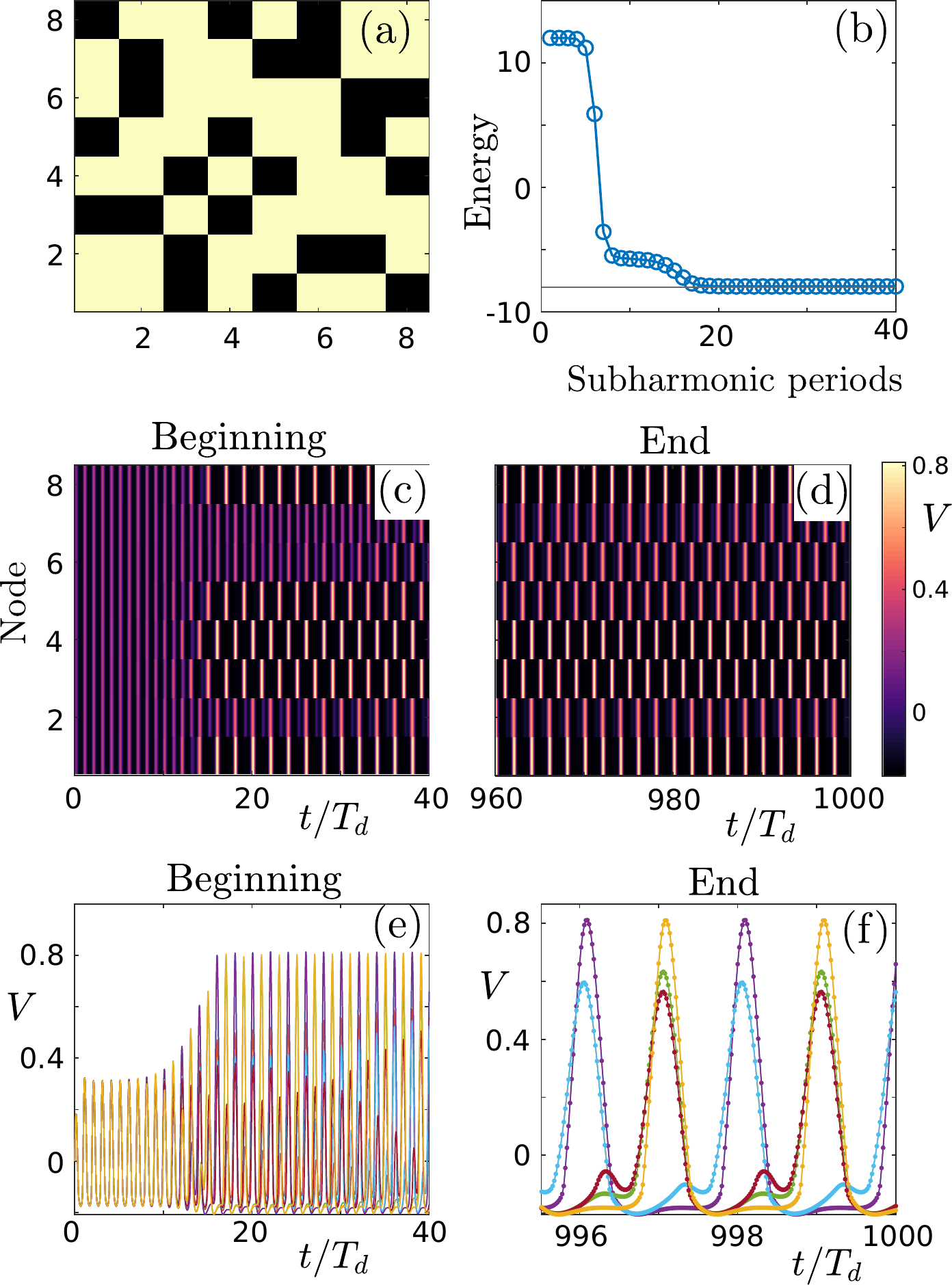}
\caption{Simulation results corresponding to the experimental setup of Fig. \ref{data1}. (a) 3-regular graph of $N=8$ coupling matrix (b) energy evolution of the network; we reach an approximate ground state of energy $E\approx-8$ after around 20 subharmonic periods. (c), (d) early and late circuit response, respectively. (e), (f) early and late time evolution of the voltage of the 8 oscillators starting from very small (${\sim}10^{-3} V$) random initial conditions (the different colors stand for different oscillators and the time unit used is the driving period $T_d$). The driving parameters used read $\omega_d=1.26\omega_0$ and $V_d=3.1 V$.}
\label{sim2}
\end{figure}

More importantly, the simulations perform well on the 3-regular graphs from before, as shown in Fig.~\ref{sim2}.
It is clear that the simulations manage to find one correct ground state of energy $E\approx-8$  within roughly 20 subharmonic periods (Fig. \ref{sim2}(b)). Figures \ref{sim2}(c) and (d) show the oscillation pattern of all 8 oscillators at an early time and at long times, respectively. The corresponding voltage traces of the oscillators are displayed in the lower two panels, (e) and (f). The same qualitative picture is observed for different initializations of the system.
It is interesting to also note how the system overcomes metastable dynamics
(i.e., between 10 and 20 subharmonic periods) to reach the desired lowest
energy minimum.
Comparing the numerical findings to the experimental results (Fig.~\ref{data1}), we see qualitative agreement in the final state and how it emerges via the establishment of the subharmonic response. For instance, in both experiment and simulation, we observe that a certain subset of oscillators moves into the subharmonic regime quickly, whereas others take significantly longer to snap into place. Furthermore, we find both experimentally and numerically that the final oscillator amplitudes are not always equal, and those oscillators that are lower in amplitude have not completely suppressed their alternate peaks and therefore exhibit a larger Fourier component at the driver frequency (Fig. \ref{sim2}(f)). The same phenomenon is apparent in Fig.~\ref{data2}, for instance, and indicates some limitations in the analogy of the electrical circuits, explored here, with Ising machines. Indeed, our oscillators are not ``true spins'' but rather are
able to feature a more complex subharmonic response in their continuous time
dependence.
One way to overcome this issue of the heterogeneity of the oscillators' amplitudes  is to introduce feedback that drives all amplitudes to the same occupation \cite{kalinin2018networks}.

The one quantitative difference that we consistently observe is that in the simulations the final state can be obtained more quickly than in the experiments. One reason for the longer times in the experimental system could be the presence of a certain level of inhomogeneity between the oscillators. Another factor could be that varactor-diode dissipation is not precisely captured in the model. Nonetheless, it is evident that the key features of the experimental results are correctly reproduced in the numerics. 

\begin{figure}
\includegraphics[width=0.98\columnwidth]{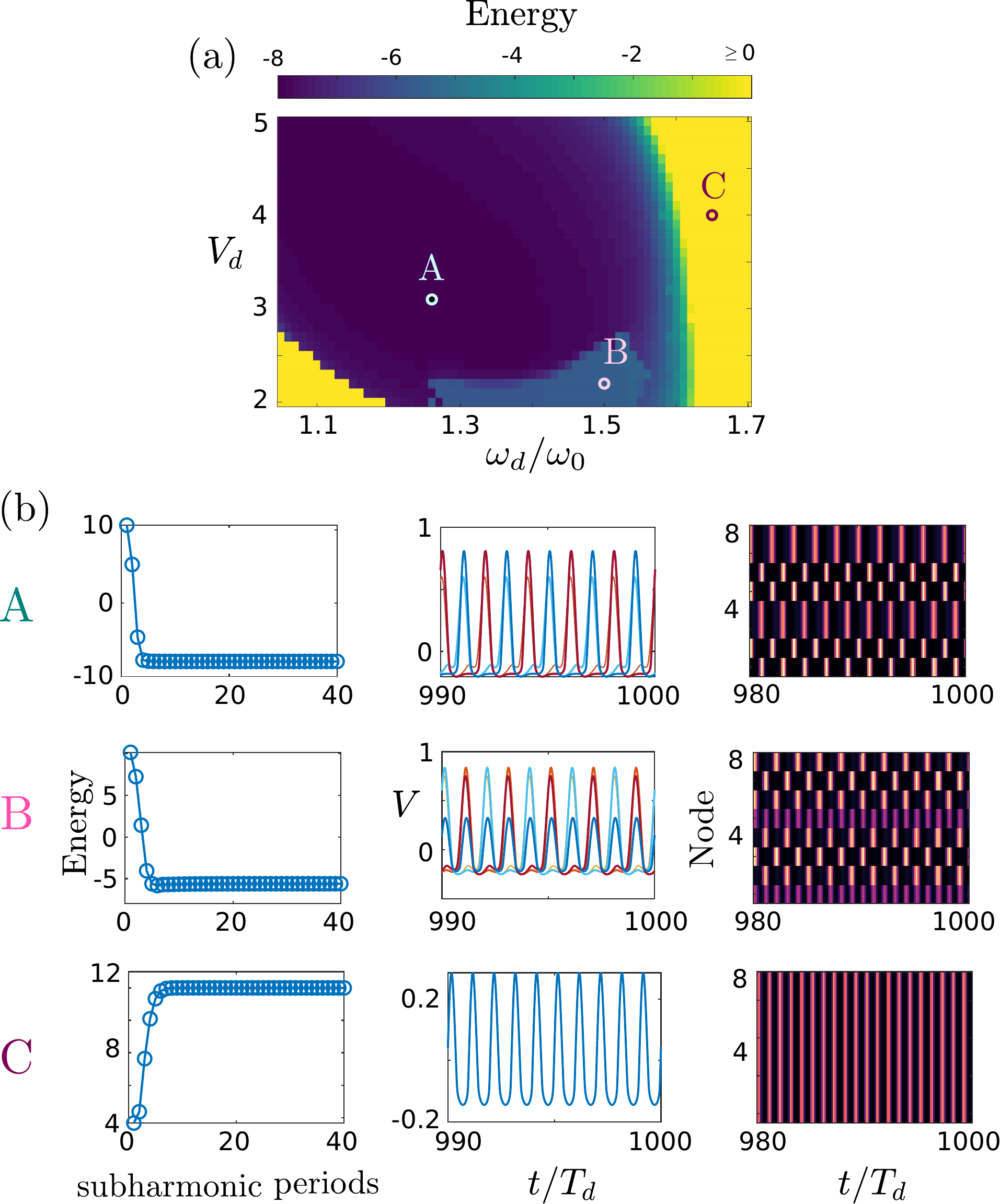}
\caption{(a) Dependence of the final energy  for the  M\"{o}bius-ladder network (Fig. \ref{data0b}) on the driving frequency $\omega_d$ and the driving amplitude $V_d$. We observe the existence of three qualitatively different regions (dark blue, blue-green, yellow), which are marked with the representative data points A ($\omega_d=1.26 \omega_0$, $V_d=3.1 V$), B ($\omega_d=1.5 \omega_0$, $V_d=2.2 V$) and C ($\omega_d=1.65 \omega_0$, $V_d=4 V$). The ensuing dynamics at the A,B,C points is shown in the three rows of (b). In particular, the first column shows the time evolution of the energy at the A,B,C points, whereas the second and third columns depict the corresponding circuit response at late times.}
\label{sim3}
\end{figure}

To explore the role of the driver (through the variation of its parameters)
in greater detail, Fig.~\ref{sim3}(a) shows the energy of the eventual state as a function of the two driving parameters - frequency $\omega_d$ ($x$-axis) and amplitude $V_d$ ($y$-axis). Evidently we can distinguish between three qualitatively different regions. The dark blue region (A) corresponds to eventual states with an energy close to the ground-state energy ($E\approx-8$) of the network in Fig.~\ref{data0b}. The oscillator response pattern (Fig.~\ref{sim3}(b), first row)  is very close to one of the
degenerate ground states, i.e., the [1, 1, -1, 1, -1, -1, 1, -1], as expected. In this region the  variation in the energy values, originates mainly from the aforementioned discrepancies on the oscillator amplitudes. 

The situation is quite different in the green-blue region (B), appearing for smaller driving amplitudes and larger driving frequencies. These parameters lead to a steady state 
with an energy  $E\approx-5.4$, in which a subset of oscillators (here 2) performs smaller amplitude oscillations with the driving frequency, while the rest performs subharmonic oscillations (Fig.~\ref{sim3}(b), second row). The subharmonic oscillations are completely lost in the yellow regions  of Fig.~\ref{sim3}(a). Note that this region includes apart from the small-frequency and small-amplitude region (where the subharmonic resonance is expected to be suppressed), also the high-frequency region with $\omega_d>1.65 \omega_0$ (C).
 For these parameter values the oscillators oscillate in phase, with the driving frequency $\omega_d$  (Fig.~\ref{sim3} (b), third row), and thus lose the desirable analogy to Ising systems. 
        
In terms of the optimal driving parameters, the experiments also show that the
optimal operating regime frequency is near the lower edge of the subharmonic resonance curve, and as the frequency increases the ground state is no longer reachable, similar to what is indicated by the region of point B in Fig.~\ref{sim3}. One difference is that in the experiment, the driver amplitude cannot be increased indefinitely. In fact, experimentally, it is advantageous to stay near the lower amplitude-threshold for subharmonic resonance. At higher amplitudes, other patterns - likely driven by inhomogeneities - become dominant. While simulation and experiment paint the same qualitative picture, differences in the details will likely become smaller with further fine-tuning of diode characteristics, especially concerning resistive dissipation. Nonetheless, it is important to stress that both experiments and current numerical simulations reach an optimal solution for 3-regular graphs, and they thus demonstrate the clear promise of this network of subharmonic LC-resonators as a purely passive unconventional computing architecture.

\section{Conclusions and Future Challenges}

In summary, we have presented a concrete experimental realization of a
nonlinear electrical oscillator circuit, operating under external drive in the
regime of subharmonic resonance and allowing for a controlled selection
of couplings, so as to realize different types of 3-regular graphs
for small number of nodes systems, such as $N=6$ and $N=8$.
We have illustrated a concrete protocol so as to interpret this
nonlinear coupled dynamical system as an effective spin-lattice
and have shown that in such an interpretation, it is possible to
reach the ground state energy, both in the case of unique minimizers
and also in the presence of frustration. The role of noise in facilitating
the departure from local minima and reaching the global minimum
has been experimentally discussed. Importantly, the understanding
of the RLC-characteristics of the relevant oscillator elements can, in principle, enable
a Kirchhoff-law based theoretical model of the system that is found
to be in very good qualitative agreement with the experimental
observations. While here we have emphasized a proof-of-principle
realization of the relevant setting, it is clear that the theoretical analysis
enables a scaling of the system to higher numbers of nodes and, as shown herein,
the consideration of both the advantages, but also the limitations of
the subharmonic oscillator response in acting as an effective spin.

As indicated also above, this experimental realization provides a useful
proof-of-principle, but also paves the way towards future efforts and associated
questions. Clearly, issues related to scalability of considerations to large $N$,
aspects related to the added wealth of phenomenology of the electrical oscillators
(in comparison to simple spin variables) and its influence on the observed
dynamics, as well as the role of noise and ensembles of realizations
(and corresponding averaging) are among the many worthwhile avenues
for further exploration. One can imagine, for instance, a large-scale implementation of this scheme that utilizes on-chip integration of the electronic circuits and coupling logic. Such studies are currently in progress and will
be reported in future publications. 


{\it Acknowledgements.} This material is based upon work supported by the US National Science Foundation under Grants DMS-1809074, PHY-2110030 (P.G.K.).

\bibliographystyle{apsrev4-1}
\bibliography{Refs}

\section*{Appendix 1: The Circuit Equations}
\label{circuiteq}
Let us think of the left oscillator in Fig.~\ref{osc} as oscillator $n$ and the right one as oscillator $m$. Let us first consider the Kirchhoff loop rule on a ``bowtie-shaped'' path; we start with the circuit point in Fig.~\ref{osc} at the bottom of the left inductor, move up across the inductor, go diagonally down (and right) across resistor $R_c(-)$, up the right inductor, and finally diagonally down (and left) across $R_c(-)$. For this closed path we can write the loop rule as,
$V_n-R_c J_{nm}+V_m-R_c J_{mn}=0$,
where $J_{nm}$ is the current through the resistor connecting the top of oscillator $n$ to the bottom of oscillator $m$, and $R_c=R_c(-)$. This implies that,
\begin{equation}
V_n+V_m=R_c (J_{nm}+J_{mn}),
\label{loop1}
\end{equation}
where we are not assuming the latter two currents to be the same.
Let is now consider another Kirchhoff loop, this time starting at the left-bottom corner of Fig.~\ref{osc}, moving up across the signal generator, down across the left capacitor, $C_d$, down further across the parallel combination of diode and inductor, and finally down across the bottom capacitor, $C_d$. Here we can write,
\begin{equation}
V_d-V_{c_1}-V_n-V_{c_2} = 0.
\label{loop2}
\end{equation}
We also know that,
\begin{equation}
C_d \frac{dV_{c_1}}{dt}=I_{+},  C_d \frac{dV_{c_2}}{dt}=I_{-}.
\label{cap}
\end{equation}
Taking the time derivative of Eq.(\ref{loop2}) and substituting Eq.(\ref{cap}), we get
\begin{equation}
\frac{d}{dt}(V_d-V_n)=\frac{1}{C_d}(I_{+}+I_{-}).
\label{interm1}
\end{equation}

Let us now consider these two currents.
$I_{+}$ is the current delivered to the $n$th oscillator via the top capacitor, and $I_{-}$ the current flowing back to the signal generator from the $n$th node. Where does this current, $I_{+}$, flow next? Part of it goes through the parallel combination of diode and inductor, and part of it becomes $J_{nm}$. Now we examine the diode more closely. It can be effectively modeled as a parallel arrangement of a nonlinear resistor with a certain current-voltage relationship, $I_D(V)$, a nonlinear capacitor $C(V)$ and a dissipation resistor $R_l$. These three will be specified in greater detail later. At present, we can therefore express $I_{+}$ as,
\begin{equation}
I_{+}=-I_D+C(V)\frac{dV_n}{dt}+\frac{V_n}{R_l}+Y_n+J_{nm},
\label{current}
\end{equation}
where $Y$ represents the current through the inductor. The minus sign is added to the first term because the diodes are oriented up in the forward direction in the circuit. It is evident that $I_{-}$ is the same as $I_{+}$ except that the last term must be replaced by $J_{mn}$.
Substituting Eq.~(\ref{current}) and its equivalent into Eq.~(\ref{interm1}), and also using Eq.~(\ref{loop1}), we arrive at:
\begin{align}
\label{interm2}
&\left[1+2\frac{C(V_n)}{C_d}\right] \frac{dV_n}{dt}=\frac{dV_d}{dt}-\frac{2}{R_l C_d}V_n + \\ \nonumber
&\frac{2}{C_d}\left[I_D(V_n)-Y_n\right]-\frac{1}{R_c C_d} (V_n+V_m) \\ \nonumber
&\frac{dY_n}{dt}=\frac{V_n}{L}.
\end{align}
We can also assume a sinusoidal driving signal, $V_d=A \sin(\omega t)$. Equation (\ref{interm2}) describes a pair of nodes, but it can be naturally generalized to a network by adding up all the coupling currents, in which case the last term of the first equation in Eq.(\ref{interm2}) would have to sum over all connected nodes $m$.
We now non-dimensionalize these governing equations by introducing $\omega_0 = 1/ \sqrt{L C_d}$ and $\tau=\omega_0 t$, as well as $v_n=V_n/A$ and $\Omega=\omega/\omega_0$. This then leads to Eq.~(\ref{govern}).

Lastly, let us cite the functional forms for $C(V)$ and $I_D(V)$ that were empirically obtained in Ref.~\cite{faustino2011}.
\begin{equation}
I_D(V)= I_s (\exp(-\beta V)-1),
\nonumber
\end{equation}
with $\beta=38.8$~V$^{-1}$ and $I_s=1.25 \times 10^{-14}$ A.
\begin{equation}
 C(V)=
\begin{cases} \begin{matrix}
C_{v}+C_{w}(V')+  C(V')^2  & \mbox{if} \quad V \leq V_{c}, \\[2.0ex]
  C_0 e^{-\alpha V} & \mbox{if} \quad V > V_{c}.
\end{matrix} \end{cases}
\nonumber
\end{equation}
Here, $V'=(V-V_c)$, $\alpha=0.456$~V$^{-1}$,
$C_{v}=C_0\exp(-\alpha V_{c})$,  $C_{w}=-\alpha C_{v}$, $C=100$~nF/V$^2$, and
$V_{c}=-0.28$~V.

\section*{Appendix 2: Configurational energy}
\label{ener}

In the context of Ising model, the energy of a N-particle spin configuration $\{S_i\}$, also known as the state of the system, is given by:
\begin{equation}
E = -\frac{1}{2} \sum_{n=1}^N \sum_{m=1}^N J_{nm} S_n S_m .
\end{equation}
In casting this coupled electrical resonator system in the form of the Ising problem, we note that there are only two stable states for our subharmonic resonators (with responses at even or odd periods of the driver), as explained previously. These are associated with spin-up and spin-down. However, transient resonator behavior can be described by superpositions of these. We associate these superpositions with angles that differ from $0$ and $\pi$; for instance, a state that is an equal superposition of the even and odd states would be reasonably associated with an angle of $\pi/2$. Thus, we keep track of each oscillator's response both at even and odd periods of the driver, $A$ and $B$ respectively, and from the ratio of these we compute an angle, $\theta_n(t)=2 \arctan(A/B)$ at each  measurement time, $t$. The energy formula then takes the form,
\begin{equation}
\label{ising}
E= -\frac{1}{2} \sum_{n=1}^N \sum_{m=1}^N J_{nm} \cos(\theta_n-\theta_m) .
\end{equation}

\end{document}